\documentclass[a4paper,10pt, twoside, twocolumn]{article}
\usepackage{multicol}
\usepackage{graphicx}
\usepackage[english]{babel}
\usepackage{natbib}
\usepackage{fancyhdr}
\usepackage{xspace}

\makeatletter
\def\fnum@figure{\footnotesize{\bfseries \figurename\nobreakspace\thefigure}}
\def\fnum@table{\footnotesize{\bfseries \tablename\nobreakspace\thetable}}
\makeatother

\makeatletter
\renewcommand\section{\@startsection {section}{1}{\z@}%
	{-3.5ex \@plus -1ex \@minus -.2ex}%
	{2.3ex \@plus.2ex}%
	{\noindent\normalsize\bfseries\uppercase}}
\makeatother

\makeatletter
\renewcommand\subsection{\@startsection{subsection}{2}{\z@}%
	{-3.25ex\@plus -1ex \@minus -.2ex}%
	{1.5ex \@plus .2ex}%
	{\noindent\normalsize\bfseries}}
\makeatother

\makeatletter
\renewcommand\subsubsection{\@startsection{subsubsection}{2}{\z@}%
	{-3.25ex\@plus -1ex \@minus -.2ex}%
	{1.5ex \@plus .2ex}%
	{\noindent\normalsize\itshape}}
\makeatother

\renewenvironment{figure}{
	\begin{oldfigure}
	\begin{center}
	}{
	\end{center}
	\end{oldfigure}}

\renewenvironment{table}{
	\begin{oldtable}
	\begin{center}
	\begin{footnotesize}
	}{
	\end{footnotesize}
	\end{center}
	\end{oldtable}}
\newenvironment{table2}{
	\begin{table*}
	\begin{center}
	\begin{footnotesize}
	}{
	\end{footnotesize}
	\end{center}
	\end{table*}}

\renewenvironment{thebibliography}[1]{
	
	\begin{oldthebibliography}{#1}
	\setlength{\itemsep}{0pt}
	\begin{small}
	}{
	\end{small}
	\end{oldthebibliography}}
	
\newcommand{\pI}{Paper~I\xspace}
\newcommand{\dd}{\ensuremath{\mathrm{d}}}

\newcommand{\reft}[1]{Table~\ref{tab:#1}}
\newcommand{\reff}[1]{Figure~\ref{fig:#1}}
\newcommand{\tn}[1]{[\emph{T.N.}: #1]}

\newcommand{\fb}[1]{{\bf{#1}}}
\newcommand{\mc}[1]{\mathcal{#1}}
\newcommand{\e}[1]{\mathrm{e}^{#1}}

\begin{document}
\thispagestyle{empty}
\onecolumn

\noindent{\LARGE\bf On the dynamical evolution of globular clusters\\ II- The isolated cluster}

\vspace{0.7cm}
\noindent{\Large Michel~H\'enon$^{\star}$, translated by Florent~Renaud$^{1,2}$}

\vspace{0.2cm}
{\footnotesize \noindent \it
$^{\star}$ Institut d'Astrophysique, Paris (now at the Observatoire de Nice)\\
$^1$ Observatoire Astronomique, Universit\'e de Strasbourg, 11 rue de l'Universit\'e, F-67000 Strasbourg, France\\
$^2$ Institute of Astronomy, University of Cambridge, Madingley Road, Cambridge, CB3 0HA, UK\\
\phantom{$^2$} \emph{florent.renaud@astro.unistra.fr}}\\

\noindent{\footnotesize Originaly received on October 23, 1964; translated on October 7, 2010}
\vspace{0.5cm}

\begin{flushright}
\begin{minipage}{12cm}
\setlength{\parindent}{15pt}

This paper is an English translation of Michel H\'enon's article, \emph{Sur l'\'evolution dynamique des amas globulaires. II- L'amas isol\'e} originally published in French in the Annales d'Astrophysique, Vol. 28, p.62 (1965).

A translation of the first paper of this series (\pI: \emph{Sur l'\'evolution dynamique des amas globulaires}, H\'enon 1961, Annales d'Astrophysique, Vol. 24, p.369) is also available.

Conventions and notations are as in the original version, for consistency. The English version is written so that it is as faithful to the French text as possible. The translator added some notes [\emph{T.N.}] for the sake of clarity, when required. French, English and Russian abstracts written by M.~H\'enon are available in the original version of the paper. The English part is reproduced below.

FR thanks Michel H\'enon for the enthusiasm and kindness he expressed when he was asked for permission to translate his work, as well as Douglas Heggie and Mark Gieles for their careful proofreading.

\end{minipage}
\end{flushright}

\vspace{0.5cm}
\begin{multicols}{2}

\section*{Original abstract}

The homologic \tn{homologous} model which was defined and studied in an earlier paper \citep{Henon1961} is computed for the case of an isolated cluster. This model is compared to von Hoerner's numerical results \citep{vonHoerner1963}. The agreement is quite good for the density low [\emph{sic} law]; the distribution of energies and the rate of escape. The broad features of the evolution are: tendency towards the homologic model, building up of an infinite central density, formation of close binaries, expansion of the cluster as a whole. An unexplained discrepancy remains in the rate of evolution.

\section{Introduction}

In a previous work \citep[hereafter \pI]{Henon1961}, we have obtained a system of fundamental equations that allows one to calculate the dynamical evolution of a star cluster from a given initial state. We have then calculated and studied in detail a particular solution of this system: the ``homologous model'', which has the property of remaining self-similar in time, the evolution being reduced to the scaling of the physical quantities.

The cluster was supposed to be plunged into an inhomogeneous field of forces, due to the Galaxy. The effect of such a field (often called ``tidal effect'') is to limit the cluster radius to a  specific value $r_e$, linked with the mass of the cluster through the relation (\pI, Equation 3.8):
\begin{equation}
r_e \propto \mc{M}_e^{1/3}.
\end{equation}

Every star further than the distance $r_e$ definitely escapes from the cluster. This way, a loss of stars is continuously happening, and the cluster mass decreases with time. From Equation~1, the radius decreases too.

The present paper focusses on the different case of an \emph{isolated cluster}, i.e. when the external forces are zero or negligible. Then, the radius $r_e$ is infinite, and the stars cannot escape anymore\footnote{A more detailed study of this phenomenon is given in \citet{Henon1960}; we find that the escape rate is not exactly zero, but negligible in practice. Similar results have been obtained by \citet{Woolley1962}.} (see \pI, Chapter~III). Hereafter, the mass $\mc{M}_e$ of the cluster is constant over time. All the equations given in \pI are still valid in this case, as soon as $\lambda = \infty$, and one can solve them by use of to numerical computations (see \pI, Chapter~V).

\end{multicols}\twocolumn\noindent 
\section{Results}

\subsection{Structure}

Once again, we find that only one unique homologous model exists. The four fundamental functions $\fb{F}(\fb{E})$, $\fb{D}(\fb{U})$, $\fb{R}(\fb{U})$, $\fb{Q}(\fb{E})$ are given in \reft{1} and plotted in \reff{1}. As in \pI, all the variables relative to the homologous model are in bold font; $\fb{E}$ is the energy, $\fb{U}$ is the potential, $\fb{F}$ is the distribution function, $\fb{D}$ the spatial density, $\fb{R}$ the distance to the centre; $\fb{Q}$ is a functional form without particular physical meaning (one can show, in fact, that $\fb{Q}$ is the \emph{adiabatic invariant} associated to the total energy $\fb{E}$, i.e. the volume of phase-space enclosed within the surface $\fb{E} = \mathrm{cst}$; see \citealt{LeviCivita1934}). \reff{1} is quite similar to the analogous figure in \pI, the main difference being that here, $\fb{R}$ goes to infinity when $\fb{U}$ goes to zero, because the cluster has an infinite radius.

\begin{table2}
\caption{The isolated homologous model} 
\label{tab:1} 
\begin{tabular}{r@{}lr@{}lr@{}lr@{}lr@{}lr@{}l}
\multicolumn{2}{c}{$\fb{E}$ or $\fb{U}$} & \multicolumn{2}{c}{$\fb{F}$} & \multicolumn{2}{c}{$\fb{D}$} & \multicolumn{2}{c}{$\fb{R}$} & \multicolumn{2}{c}{$\fb{Q}$} & \multicolumn{2}{c}{$\fb{D}_p$} \\
\hline
\hline
-5&   & 131&.81 & 134&.66 & 0&.1126 & 0&.0002927 & 39&.86\\
-4&.8 & 106&.43 & 106&.45 & 0&.1256 & & \phantom{.000}4025 & 34&.72\\
-4&.6 & 85&.80 & 83&.83 & 0&.1402 & & \phantom{.000}5554 & 30&.13\\
-4&.4 & 69&.05 & 65&.71 & 0&.1566 & & \phantom{.000}7680 & 26&.05\\
-4&.2 & 55&.46 & 51&.26 & 0&.1753 & 0&.001064 & 22&.41\\
-4&   & 44&.44 & 39&.75 & 0&.1964 & & \phantom{.00}1479 & 19&.18\\
-3&.8 & 35&.52 & 30&.63 & 0&.2204 & & \phantom{.00}2060 & 16&.33\\
-3&.6 & 28&.30 & 23&.43 & 0&.2478 & & \phantom{.00}2879 & 13&.80\\
-3&.4 & 22&.47 & 17&.76 & 0&.2793 & & \phantom{.00}4039 & 11&.58\\
-3&.2 & 17&.77 & 13&.34 & 0&.3153 & & \phantom{.00}5691 & 9&.629\\
-3&   & 13&.98 & 9&.898 & 0&.3571 & & \phantom{.00}8058 & 7&.928\\
-2&.8 & 10&.94 & 7&.247 & 0&.4058 & 0&.01148 & 6&.451\\
-2&.6 & 8&.495 & 5&.220 & 0&.4629 & & \phantom{.0}1647 & 5&.176\\
-2&.4 & 6&.544 & 3&.689 & 0&.5304 & & \phantom{.0}2382 & 4&.085\\
-2&.2 & 4&.989 & 2&.546 & 0&.6110 & & \phantom{.0}3479 & 3&.160\\
-2&   & 3&.754 & 1&.707 & 0&.7085 & & \phantom{.0}5141 & 2&.384\\
-1&.8 & 2&.777 & 1&.104 & 0&.8283 & & \phantom{.0}7709 & 1&.744\\
-1&.6 & 2&.010 & 0&.6818 & 0&.9782 & 0&.1177 & 1&.226\\
-1&.4 & 1&.413 & 0&.3958 & 1&.171 & & \phantom{.}1837 & 0&.8170\\
-1&.2 & 0&.9516 & 0&.2112 & 1&.425 & & \phantom{.}2955 & 0&.5060\\
\\
-1&   & 0&.6018 & 0&.09978 & 1&.778 & & \phantom{.}4948 & 0&.2816\\
-0&.9 & 0&.4623 & 0&.06429 & 2&.011 & & \phantom{.}6533 & \\
-0&.8 & 0&.3437 & 0&.03904 & 2&.300 & & \phantom{.}8779 & 0&.1324\\
-0&.7 & 0&.2444 & 0&.02191 & 2&.667 & 1&.206 & \\
-0&.6 & 0&.1631 & 0&.01103 & 3&.151 & 1&.706 & 0&.0462\\
-0&.5 & 0&.09898 & 0&.004740 & 3&.822 & 2&.510 & 0&.0227\\
-0&.4 & 0&.05150 & 0&.001584 & 4&.817 & 3&.909 & 0&.00928\\
-0&.3 & 0&.02015 & 0&.000350 & 6&.457 & 6&.66 & 0&.00262\\
-0&.2 & 0&.00371 & 0&.000033 & 9&.710 & 13&.4: & 0&.000346\\
-0&.1 & 0&.00011 & 0&.000000 & 19&.41 & 41&.1: & 0&.000005\\
 0&   & 0&       & 0&        & &$\infty$ && $\infty$ & 0\\
\hline
\end{tabular}
\end{table2}

\begin{figure}
\includegraphics[width=\columnwidth]{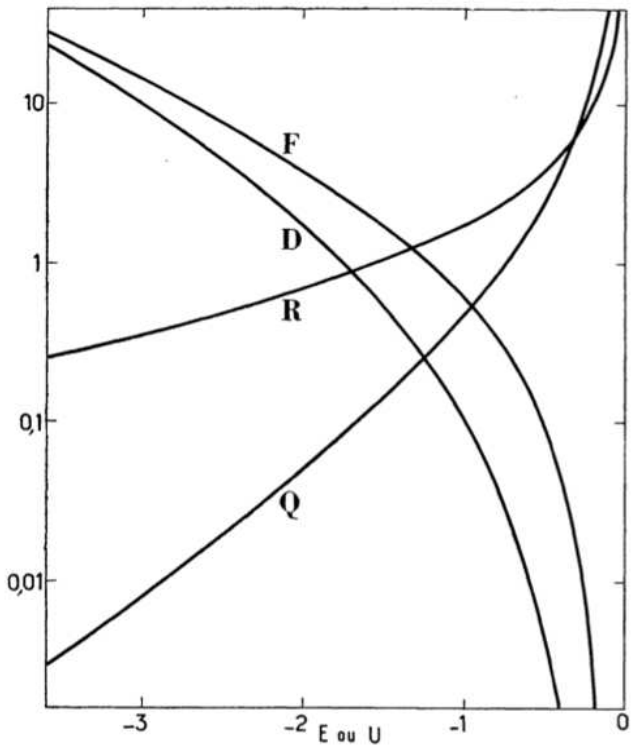}
\caption{Isolated homologous model: the four fundamental functions $\fb{F}(\fb{E})$, the distribution function;  $\fb{D}(\fb{U})$, the spatial density; $\fb{R}(\fb{U})$, the distance to the centre; $\fb{Q}(\fb{E})$, the adiabatic invariant.}
\label{fig:1}
\end{figure}

Below $\fb{E} = -5$ or $\fb{U} = -5$, one can use the asymptotic forms of (4.34) and (4.35) from \pI. Furthermore, for $\fb{E}\to 0$, or $\fb{U}\to 0$, the asymptotic forms are:
\begin{eqnarray}
\fb{F} &=& a_0\ \e{-\sigma'\sqrt{-\fb{E}}},\\
\fb{D} &=& 2 \sqrt{\pi} a_0\ \sigma^{-3/2} (-\fb{U})^{9/4}\ \e{-\sigma'\sqrt{-\fb{U}}},\nonumber\\
\fb{R} &=& -\fb{M}_e / \fb{U},\nonumber\\
\fb{Q} &=& \frac{\pi\ \fb{M}_e^3}{12\sqrt{2}}(-\fb{E})^{-3/2},\nonumber
\end{eqnarray}
with:
\[ a_0 = 6.391, \qquad \sigma = 3.544. \]
The values of constants found in the equations are:
\begin{eqnarray}
K & = & -1.363,\\
c & = & -1.097,\nonumber\\
b & = & +\frac{2}{3}c = -0.7313,\nonumber\\
\fb{M}_e & = & 1.946 \qquad \textrm{(total mass)},\nonumber\\
\fb{L}_e & = & 1.217 \qquad \textrm{(total kinetic energy)}.\nonumber
\end{eqnarray}

The spatial density $\fb{D}$ is plotted in \reff{2} as a function of the radius, and compared to those of the non-isolated model (the two curves are normalized to get the same asymptotical form for $\fb{R}\to 0$). Toward the outskirts, the models are naturally very different, as one yields a finite radius while the other is infinite. However, close to the centre, both models overlap. Therefore, the presence of an external field only modifies the outer layers of the cluster, as one could have expected. To be clearer, we can state that the two models are almost identical within the sphere enclosing half of the total mass (the radius of this sphere, the so-called median (\tn{half-mass}) radius, is $\fb{R} = 0.7563$ for the isolated cluster). In particular, we note that the isolated cluster, like the non-isolated cluster, yields a density that goes to infinity as $\fb{R}^{-2}$ when $\fb{R}$ goes to zero.

\begin{figure}
\includegraphics[width=\columnwidth]{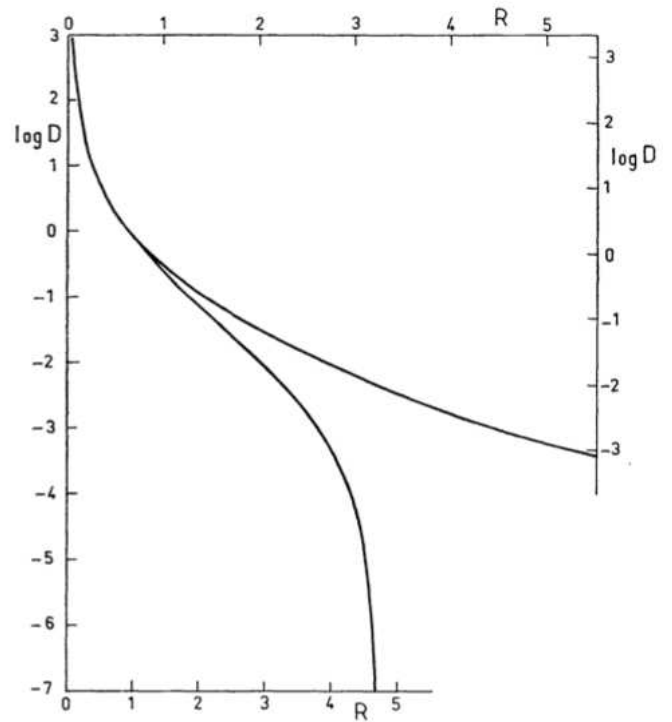}
\caption{spatial density of the isolated homologous model (on top and right) and of the non-isolated model (on bottom and left), as a function of the radius.}
\label{fig:2}
\end{figure}

The last column of \reft{1} gives the projected density $\fb{D}_p$ as a function of $\fb{R}$. It goes to infinity as $\fb{R}^{-1}$ when $\fb{R}$ goes to zero. For $\fb{R} \to \infty$, its asymptotic form is:
\begin{equation}
\fb{D}_p = 4\pi a_0\ \sigma^{-2}\ \fb{M}_e^{5/2}\ \fb{R}^{-3/2} \e{-\sigma \sqrt{\fb{R}/\fb{M}_e}}.
\end{equation}

\subsection{Evolution}

The equations (5.14) of \pI, which give the two parameters $\beta$ and $\gamma$ of the homology as functions of time, become:
\begin{eqnarray}
\beta & = & \beta_0 \left(1 - \frac{T}{T_1}\right)^{-2/3},\\
\gamma & = & \gamma_0 \left(1 - \frac{T}{T_1}\right)^{-1}.\nonumber
\end{eqnarray}
From this, one can derive the variations of the physical quantities as functions of time. In particular, the radius of the cluster evolves as
\begin{equation}
r = r_0 \left(1 - \frac{t}{t_1}\right)^{2/3}.
\end{equation}
Here, $r$ represents any quantity which characterizes the size of the cluster, e.g. the median radius defined above. $t_1$ is negative (see Paper I, 5.13). The variation of $r$ is plotted on \reff{3}. We see that the isolated cluster \emph{expands} with time (while the non-isolated cluster contracts).

\begin{figure}
\includegraphics[width=\columnwidth]{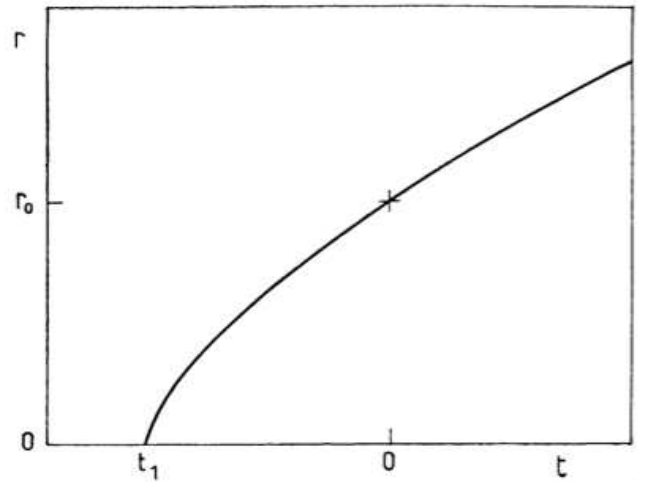}
\caption{Isolated homologous model: evolution of the radius.}
\label{fig:3}
\end{figure}

This result differs from the usual idea according to which internal encounters make the cluster contract. This can be easily explained. The classical reasoning is as follows: the stars which escape from the cluster take with them practically zero energy, so that the total energy of the cluster remains constant; according to the virial theorem, the potential energy also remains constant; as the total number of stars decreases, the distances between them must decrease too. But here, on the one hand there are no escapers because the cluster is isolated. On the other hand, the fact that the parameter $K$ is non-zero and negative (as shown in \pI, Chapter~IV) implies the existence of a \emph{negative flux of energy} toward the centre of the cluster. The phenomenon occurs in the same way as for non-isolated clusters. This negative energy accumulates in the centre, thanks to the formation of increasingly close binaries or multiple stars. At the same time, the energy in the rest of the cluster decreases (in absolute value); inducing an increase of the distance between the stars, i.e. a dilation of the cluster.

The curve on \reff{3} continues to infinity in the future; it is therefore impossible to define a remaining lifetime for the cluster, as opposed to the non-isolated case. In theory, the evolution of the cluster goes on indefinitely. In practice however, while the cluster expands continuously, it will eventually feel the gravitational effect of another object; we are back to the non-isolated case studied in \pI.

The curve in \reff{3} is limited toward small $t$'s. Thus, one can obtain an upper limit equal to $|t_1|$ for the age of the isolated cluster. This is strictly true only when one assumes that the cluster has always followed the homologous model, but is probably true in an approximate fashion in other cases. This limit can be calculated for a real cluster of which one knows the mass and size, thanks to (2.24), (5.13) and (8.18) of \pI; one obtains:
\begin{equation}
|t_1| = \frac{1}{G^{1/2}\ m\ |c|\ \ln{(n)}} \left(\frac{\mc{M}_e}{\fb{M}_e}\right)^{1/2} \left(\frac{r}{\fb{R}}\right)^{3/2}.
\end{equation}

\section{Comparisons with the numerical experiments of von Hoerner}

There is no such thing as a real system with which the present model could be usefully compared. The open clusters and the globular clusters can surely not be considered as isolated; their observed radius is close to the limit set by the galactic field, in general. The elliptical galaxies could, in many cases, be considered as isolated systems to a good approximation; but their internal relaxation time is extremely long, so that their structure is probably not set by \tn{stellar} encounter effects. Finally, for galaxy clusters, the distribution function cannot be precisely determined, in particular in the external regions.

However, the present model can be compared with the two ``experimental clusters'' of \citet[hereafter v.H.]{vonHoerner1963}. These are imaginary clusters made of 25 stars, whose evolution has been numerically computed for a time much longer than the relaxation time. No other forces than mutual attraction are taken into account; therefore, these clusters lie in the isolated case. Furthermore, the masses of the stars are taken to be equal; as in the homologous model.

\subsection{Density law}

The evolution of the spatial distribution of the clusters is plotted in v.H., Figure~5. For the comparison, we consider the final state, the most relaxed, i.e. $t=5.4$ in the first (\tn{experimental cluster}) case and $t=2.9$ in the second case (corresponding to the mean of the fifth and fourth last columns of Table~2 of v.H., respectively). In order to compare these distributions with those of the homologous model, it is necessary to scale $r$ and $\rho$ correctly. These scalings derive from the median radius and the total mass. The two clusters have median radii of 0.840 and 1.006 respectively (from v.H., Table~10) and a total mass of unity; the homologous model has a median radius of 0.7563 and a mass of:
\begin{equation}
\int_0^{\infty} 4\pi \fb{D}\ \fb{R}^2\, \dd\fb{R} = 4\pi \fb{M}_e = 24.45 \qquad \textrm{(\pI, 2.45)}.
\end{equation}
One finds:
\begin{eqnarray}
\fb{R}/r & = & 0.900 \quad \textrm{or} \quad 0.752;\\
\fb{D}/\rho & = & 33.5 \quad \textrm{or} \quad 57.5.\nonumber
\end{eqnarray}

\reff{4} shows the result. The agreement \tn{between the theoretical and numerical relations} is quite good, especially when considering that there is no free parameter in this comparison, and when noticing that the initial state of the clusters, also visible on \reff{4} was very different from that of the homologous model.

\begin{figure}
\includegraphics[width=\columnwidth]{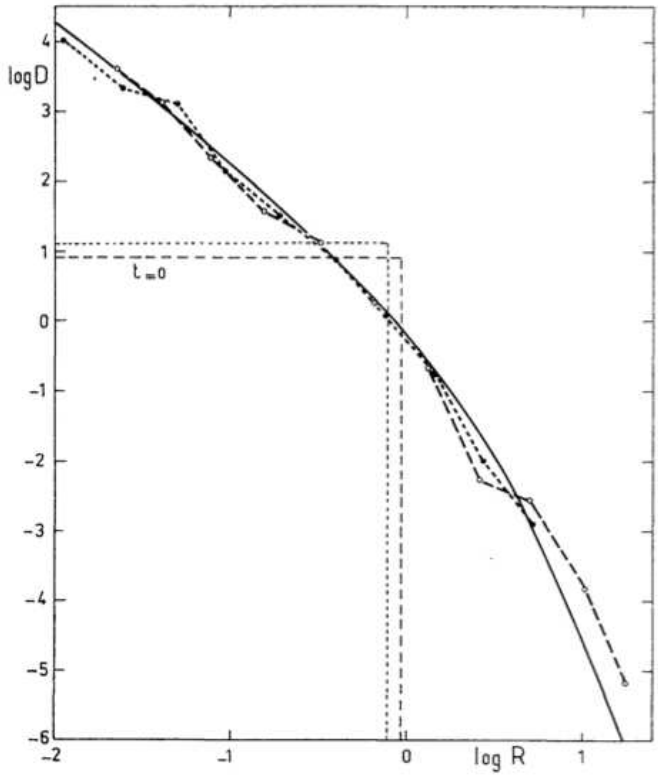}
\caption{Comparison of the density law: homologous model (solid line) and numerical experiments (\tn{dotted and} dashed lines).}
\label{fig:4}
\end{figure}

\subsection{Energy distribution}

von Hoerner also gives the energy distribution of the stars (v.H., Figure~9 and last column of Table~14). The energy is normalized to its mean value $\overline{\fb{E}}$, and transformed into another quantity $\varepsilon$, for graphical purposes:
\begin{equation}
\varepsilon = \textrm{arcsinh}\left(\frac{\fb{E}}{\overline{\fb{E}}}\right);
\end{equation}
and $f(\varepsilon)\ \dd\varepsilon$ is the fraction of the stars whose energy is within $\varepsilon$ and $\varepsilon + \dd\varepsilon$.

To obtain the corresponding curve in the case of the homologous model, one must first calculate the mean energy of a star. It is easy to demonstrate that, for a virialised cluster, this mean energy is minus three times the mean kinetic energy (see v.H., Eq.~23), i.e.:
\begin{equation}
\overline{\fb{E}} = -3 \frac{\fb{L}_e}{\fb{M}_e} = -1.876.
\end{equation}

Furthermore, the fraction of the stars having an energy between $\fb{E}$ and $\fb{E} + \dd\fb{E}$ is (\pI, 2.42):
\[\frac{\fb{F} \fb{Q}'\ \dd\fb{E}}{\fb{M}_e}.\]
When using the new variable $\varepsilon$, this expression must be equal to $f(\varepsilon)\ \dd\varepsilon$. Thus, using (10), one gets:
\begin{equation}
f(\varepsilon) = \frac{\fb{F} \fb{Q}'\ \sqrt{\fb{E}^2+\overline{\fb{E}}^2}}{\fb{M}_e}.
\end{equation}

The distribution obtained is plotted in \reff{5}, and is compared with the results of von Hoerner. Note that, here again, there is no free-parameter. The agreement is excellent from $\varepsilon = 0$ to $\varepsilon = 2.5$, i.e. for the energies ranging from 0 to about 6 times the mean energy. This interval contains almost all the stars (see v.H., Figure~8). In particular, the ``shoulder'' of the distribution next to $\varepsilon = 0$ is very well reproduced. For $\fb{E}$ [\emph{sic}, $\varepsilon$] larger than 2.5, the results differ; the stars with highly negative energies are in excess \tn{in the numerical experiments of von Hoerner} compared to the homologous model. This probably comes from the formation of very close binaries, detected by von Hoerner (his Table~5), and predicted by theory as described above.

\begin{figure}
\includegraphics[width=\columnwidth]{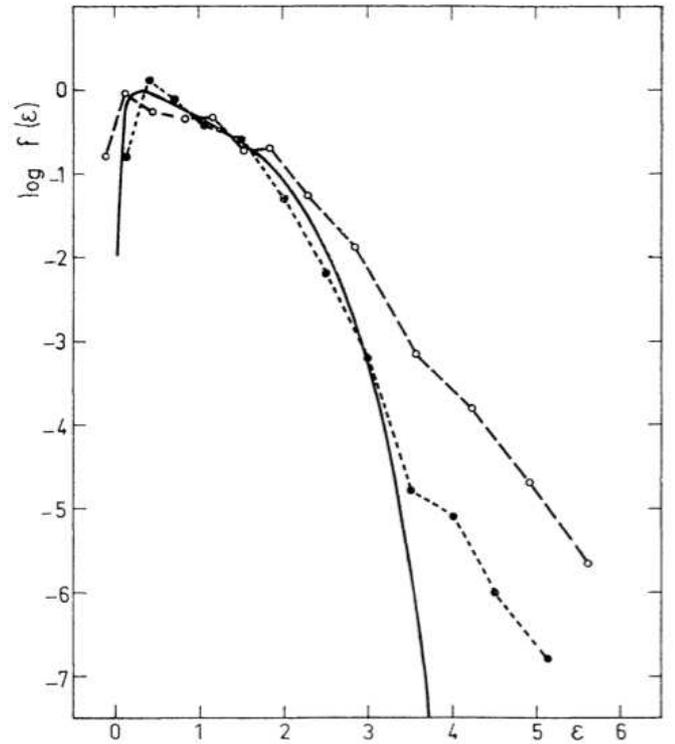}
\caption{Comparison of the energy distribution: homologous model (solid line) and numerical experiments (\tn{dotted and} dashed lines).}
\label{fig:5}
\end{figure}

Let's digress about this topic. One could be surprised that the theory leads to the disagreement illustrated in \reff{5}, although it predicted the existence of binaries. In fact, the homologous model presented here, as well as all the other theoretical models presented up to now, to our knowledge, is fundamentally unable to account for binaries. We assume, indeed, that the state of the cluster is fully described by the distribution function (\pI, 2.1):
\[\varphi(x,y,z,v_x,v_y,v_z,m,t) \quad \textrm{or, in short:}\quad \varphi(\tau,t),\]
which gives the probability to have a star in a small volume of phase-space, at a given time. Doing this, we implicitly assume that the stars are \emph{independent} from each other, meaning that the probability of finding a star at a given point of phase-space does not depend on the presence or absence of stars in other points of phase-space. However, this is only a hypothesis and it is not necessarily true. It seems that this fundamental point has not received all the care it deserves; it is even ignored in the classical stellar dynamics studies.

But this hypothesis contradicts the existence of binaries in the cluster. Indeed, if pairs exist, the presence of a star at a point of (ordinary) space leads to a higher probability to find another star in the neighbourhood. In phase-space, the probability is also modified. To get a correct mathematical description of this effect, we must introduce the \emph{double distribution function}:
\[\varphi_2(\tau_1, \tau_2, t),\]
which gives the probability that a star is at the point $\tau_1$ \emph{and} another star is at the point $\tau_2$, at the time $t$.

The same way, it would be necessary to introduce a triple distribution function $\varphi_3$ to account for triple stars, and so on. These multiple distribution functions are common in plasma physics \citep[see e.g.][Chapter~6]{Delcroix1963} and it would be interesting to introduce them in stellar dynamics too. This would perhaps allow us to calculate exactly, e.g., the formation rate and the properties of binaries in a cluster. We could also expect from this a more precise theory about the effect of encounters.

\subsection{Evolution}

In \pI, Chapter~VII, we have shown that the homologous model is the final state toward which a cluster tends, whatever its initial form was. This is clearly confirmed by the results of von Hoerner (v.H., Figures~5 and 6): his clusters, which initially have a constant density inside a sphere, progressively concentrate in the centre, and form a very extended halo at the same time; the density law tends toward the theoretical curve of the homologous model (\reff{4}).

We have also shown that in a cluster which has initially a finite central density, this density rapidly increases and tends toward infinity. This phenomenon is clearly visible in the Figure~4 of von Hoener, where the ``density radius'' $\mathrm{R}_\rho$ tends to zero. It is easy to verify (v.H., Eq.~10) that this radius can become zero only if the central density is infinite. One can also directly see the rapid growth of the central density (v.H., Table~9).

Finally, we have seen above that the isolated homologous model expands with time (\reff{3}). This also appears on the right-hand side of the Figure~6 of von Hoerner: once the clusters have reached the homologous state, the evolution is only a dilatation of the entire cluster, which is seen on the figure as a translation of all the points to the top. This is especially visible in the second case.

Now, let's make a quantitative comparison. We first have to adapt the timescales. We have (\pI, Eq.~8.9):
\begin{equation}
\frac{\dd t}{\dd\fb{T}} = G^{-1/2}\ m^{-1}\ [\ln{(n)}]^{-1} \left(\frac{\mc{M}_e}{\fb{M}_e}\right)^{1/2} \left(\frac{r}{\fb{R}}\right)^{3/2};
\end{equation}
$\fb{T}$ being the time variable of the homologous model, while $t$ is the real time. We have $\mc{M}_e = mn$; and when using the definition of the unit of time given by von Hoerner (v.H., p.54), we get:
\begin{equation}
\frac{\dd t}{\dd\fb{T}} = n^{1/2}\ [\ln{(n)}]^{-1}\ \fb{M}_e^{-1/2} \left(\frac{r}{\fb{R}}\right)^{3/2}.
\end{equation}
The $r/\fb{R}$ ratio does not vary much with time (v.H., Table~10); we can assume it remains constant and adopt the values given by (9). With $n=25$, we get:
\begin{equation}
\frac{\dd t}{\dd\fb{T}} = 1.303 \quad \textrm{or} \quad 1.707.
\end{equation}

From \pI, (2.32) and (5.13), we find out:
\begin{equation}
t_1 = -1.188 \quad \textrm{or}\quad -1.556,
\end{equation}
and (6) gives the theoretical rate of change of the radius, at $t=0$:
\begin{equation}
\frac{\dd\log{(r)}}{\dd t} = 0.244 \quad \textrm{or}\quad 0.186.
\end{equation}

The observed rate of change for the median radius is (v.H., Table~10; we used the mean value, over the entire evolution):
\begin{equation}
\frac{\dd\log{(r_\mathrm{med})}}{\dd t} = 0.026 \quad \textrm{or}\quad 0.031.
\end{equation}
Here, we find a clear disagreement: the two clusters evolve much slower than predicted by theory, by a factor of 10 or 6, approximatively.

We can also calculate the time required to form an infinite central density. Back to the calculation made in \pI, Chapter~VII, and assuming that the parameter $d$ takes the same value here, we find that a isolated cluster, starting from a weakly concentrated state, must get an infinite central density after a time:
\begin{equation}
\fb{T}_2= 0.3052,
\end{equation}
i.e.
\begin{equation}
t_2 = 0.473 \quad \textrm{or}\quad 0.620,
\end{equation}
from \pI, (5.12), (2.32) and (15) above.

But the Figure~4 of von Hoerner shows that the infinite central density appears after about $t = 6$ for the first cluster, and $t = 4$ for the second. Here again, the observed evolution rate is smaller than the theoretical one, by a factor 13 or 6.

The reason for such a disagreement has not been found.

\subsection{Escape rate}
As mentioned above, the escape rate of an isolated cluster is much smaller than that of a non-isolated cluster, but is not exactly zero. It reads \citep[Eq.~15]{Henon1960}:
\begin{equation}
\frac{\dd n}{\dd t} = -0.00868 \sqrt{\frac{Gmn}{r_0^3}},
\end{equation}
where $r_0$ is the radius enclosing, in projection, the half of the total mass.

The clusters from von Hoerner have initially the shape of an homogeneous sphere of radius unity; in this case, we have $r_0 = 0.6083$. Furthermore, $n=25$. Taking into account the choice for the unit of time (\tn{v.H.,} p. 54), we obtain:
\begin{equation}
\frac{\dd n}{\dd t} = -0.0915.
\end{equation}
The total duration of the two examples calculated with $n=25$ is (v.H., Table~8):
\[\Delta t = 6.23 + 3.86 = 10.09;\]
so that the number of escapers should be $\Delta n = -0.92$. In fact, we measure: $\Delta n = -1$ (v.H., Table~6).

Four other examples have been calculated with $n=16$. Their duration is not given by von Hoerner. However, the total number of relaxation times covered is 88, instead of 64 for the examples with $n=25$. The number of escapers should be larger in proportion, i.e. $\Delta n = -1.27$. The number measured is: $\Delta n = -1$. We note that here, the agreement between the theory and the numerical experiments is very good.

\section{Conclusion}

We have compared the results obtained for the same problem: structure and evolution of an isolated cluster, thanks to two very different methods, almost opposite: the first one (the homologous model) describes the cluster with a continuous distribution function, whose evolution is ruled by a Fokker-Planck type equation; the second one (numerical experiments) represents the cluster as a discrete system made of material points, whose evolution is given by the ordinary laws of mechanics.

The results are generally in very good agreement. This constitutes a solid argument in favor of the validity of both theoretical methods.

Some hypotheses are also confirmed. In the case of the homologous model, we had to assume an isotropic distribution for the velocities everywhere in space, in order to reduce the complexity of the equations. No restriction of this kind exists in the numerical experiments, where the entire evolution of the system of points is computed exactly. The agreement of the results, particularly for the density law (\reff{4}), indicates that this hypothesis about isotropy does not strongly affect the structure of the cluster. We can reasonably suppose that the same applies to the non-isolated clusters, studied in \pI.

Furthermore, in the case of the numerical experiments, one could ask whether a system of only 25 points can properly represent a real cluster, made of a much larger number of stars. But the homologous model, with its continuous distribution function, can be considered as the limiting case of a system made of an infinite number of points. The similar results obtained in both cases are thus, very probably, also valid for all the intermediate cases, i.e. from $n=25$ to $n=\infty$; this range includes all the real stellar systems, which are thus ``surrounded'' by the two methods. We can also claim that, for the questions we address, 25 is a good approximation of infinity.

Therefore, it seems recommended to develop further the method of the numerical experiments, which is still in its early stages. Addressing simultaneously the questions of stellar dynamics with theory and numerical experiments would, certainly, allow us to make more solid and faster progresses than in the past.


\end{document}